\documentclass[a4paper,fleqn,usenatbib]{mnras}

\usepackage{newtxtext,newtxmath}

\usepackage[T1]{fontenc}
\usepackage{ae,aecompl}

\usepackage{graphicx}	
\usepackage{amsmath}	
\usepackage{amssymb}	
\usepackage{siunitx}

\newcommand{\lya}{Ly$\alpha$}

\newcommand{\sbcgs}{$\rm erg~s^{-1}~cm^{-2}~$\AA$^{-1}~\rm arcsec^{-2}$}
\newcommand{\sbline}{$\rm erg~s^{-1}~cm^{-2}~arcsec^{-2}$}

\newcommand{\qsoone}{J214225.78$-$442018.3}
\newcommand{\qsotwo}{J214222.17$-$441929.8}

\newcommand{\zn}{$z_{\rm neb\, Ly\alpha}$}

\newcommand{\rev}[1]{{ #1}}
\newcommand{\revs}[1]{{ #1}}
\newcommand{\revt}[1]{{ #1}}

\DeclareRobustCommand{\ion}[2]{%
\relax\ifmmode
\ifx\testbx\f@series
{\mathbf{#1\,\mathsc{#2}}}\else
{\mathrm{#1\,\mathsc{#2}}}\fi
\else\textup{#1\,{\mdseries\textsc{#2}}}%
\fi}

\title[MUDF I. Nebulae in a  $z\approx 3.23$ quasar pair]{The MUSE Ultra Deep Field (MUDF). I. Discovery of a group of Ly$\alpha$ nebulae associated with a bright $z\approx 3.23$ quasar pair}

\author[E. Lusso et al.]{E. Lusso$^{1}$\thanks{E-mail: elisabeta.lusso@durham.ac.uk},
M. Fumagalli$^{1,2}$,
M. Fossati$^{1,2}$, 
R. Mackenzie$^{3}$,
R. M. Bielby$^{1}$, \newauthor
F. Arrigoni Battaia$^{4,5}$,
S. Cantalupo$^{3}$,
R. Cooke$^{1,2}$,
S. Cristiani$^{6}$,
P. Dayal$^{7}$,\newauthor
V. D'Odorico$^{6,8}$,
F. Haardt$^{9}$,
E. Lofthouse$^{1,2}$, 
S. Morris$^{1,2}$,
C. Peroux$^{10}$,\newauthor
L. Prichard$^{11}$, 
M. Rafelski$^{11,12}$,
R. Simcoe$^{13}$,
A. M. Swinbank$^{1}$,
and T. Theuns$^{2}$.
\\
$^{1}$Centre for Extragalactic Astronomy, Durham University, South Road, Durham, DH1 3LE, UK\\
$^{2}$Institute for Computational Cosmology, South Road, Durham DH1 3LE, UK\\
$^{3}$Department of Physics, ETH Zurich, Wolfgang-Pauli-Strasse 27, 8093 Zurich, Switzerland\\
$^{4}$European Southern Observatory, Karl-Schwarzschild-Str. 2, D-85748, Garching bei Munchen, Germany\\
$^{5}$Max-Planck-Institut fur Astrophysik, Karl-Schwarzschild-Str 1, D-85748 Garching, Germany\\
$^{6}$INAF-Astronomical Observatory, via Tiepolo 11, I-34143 Trieste, Italy\\
$^{7}$Kapteyn Astronomical Institute, Rijksuniversiteit Groningen, Landleven 12, Groningen, 9717 AD, The Netherlands\\
$^{8}$Scuola Normale Superiore, P.zza dei Cavalieri, 7 I-56126 Pisa\\
$^{9}$Dipartimento di Scienza e Alta Tecnologia, Universit\`a dell'Insubria, Via Valleggio 11, I-22100 Como, Italy\\
$^{10}$Aix Marseille Universit\'e, CNRS, LAM (Laboratoire d'Astrophysique de Marseille) UMR 7326, 13388, Marseille, France\\
$^{11}$Space Telescope Science Institute, 3700 San Martin Drive, Baltimore, MD 21218, USA\\
$^{12}$Department of Physics \& Astronomy, Johns Hopkins University, Baltimore, MD 21218, USA\\
$^{13}$MIT-Kavli Institute for Astrophysics and Space Research, 77 Massachusetts Ave. 37-664D, Cambridge, MA 02139, USA\\
}

\pubyear{2018}

\begin{document}
\label{firstpage}
\pagerange{\pageref{firstpage}--\pageref{lastpage}}
\maketitle

\begin{abstract}
We present first results from Multi Unit Spectroscopic Explorer (MUSE) observations at the Very Large Telescope in the MUSE Ultra Deep Field (MUDF), a $\approx 1.2\times 1.4$ arcmin$^2$ region for which we are collecting $\approx$200 hours of integral field spectroscopy. The $\approx 40$-hour observation completed to date reveals the presence of a group of three Ly$\alpha$ nebulae associated with a bright quasar pair at $z\simeq3.23$ with projected separation of $\approx 500\rm~kpc$. 
Two of the nebulae are physically associated with the quasars which are likely powering the Ly$\alpha$ emission, and extend for $\gtrsim 100~\rm kpc$ at a surface brightness level of $\approx 6\times 10^{-19}$~\sbline. A third smaller ($\approx$35 kpc) nebula lies at a velocity offset of $\approx 1550$ km s$^{-1}$. Despite their clustered nature, the two large nebulae have properties similar to those observed in isolated quasars and exhibit no sharp decline in flux at the current depth, suggesting an even more extended distribution of gas around the quasars. \revt{We interpret the shape and the alignment of the two brighter nebulae as suggestive of the presence of an extended structure connecting the two quasar host galaxies}, as seen for massive galaxies forming within gas-rich filaments in cosmological simulations.
\end{abstract}




\section{Introduction}

In the current Cold Dark Matter (CDM) cosmological paradigm, galaxies form in overdense regions of the universe where cold ($\approx 10^4$K) gas is accreted from the intergalactic medium (IGM) into the dark matter haloes to form stars (e.g. see \citealt{2018PhR...780....1D} for a recent review). As halos assemble at the intersection of baryon-rich dark matter filaments, theory predicts the emergence of a ``cosmic web'' connecting galaxies on scales of megaparsecs \citep{bond1996}.
In the proximity of galaxies, on scales of tens to hundreds of kiloparsecs from the halo centres, these filaments are predicted to feed the circumgalactic medium (CGM), the gaseous component that is responsible for regulating the gas exchange between galaxies and the surrounding IGM \citep{tumlinson17}. 
This continuous refuelling of gas from filaments through the CGM, together with the ejection of baryons from halos due to feedback processes, is believed to be the key element that regulates the growth of galaxies through cosmic time. 


On the observational side, this theoretical framework has been explored over the past decades using absorption-line spectroscopy, which provides a powerful way to map the low-density IGM and CGM around galaxies \citep[e.g.][]{steidel10}. More recently, an effective technique to map the gas distribution in the CGM has been through the direct imaging of the fluorescent \lya\ line in emission around  bright quasars (e.g. \citealt{cantalupo2014,hennawi2015,borisova2016,2018MNRAS.473.3907A}) and galaxies \citep{Leclercq2017,wisotzki2018}.
Instrumental to these advances has been the deployment of the Multi Unit Spectroscopic Explorer  (MUSE; \citealt{bacon2010}) on the Very Large Telescope (VLT). 

Building on this advancement, we started a new Large Programme (ID 1100.A$-$0528; PI Fumagalli) to collect $\approx 200$ hours of MUSE observations in a $\approx 1.2\times 1.4$ arcmin$^2$ region centred at $21^h$:$42^m$:$24^s$ $-44^\circ$:$19^m$:$48^s$. 
Additional {\it Hubble Space Telescope} WFC3/G141 slitless spectroscopy in the near infrared will be collected in cycle 26 for a total of 90 orbits (PID 15637; PIs Fumagalli and Rafelski). This field, dubbed the MUSE Ultra Deep Field (MUDF), hosts several astrophysical structures at different redshifts, including two physically associated quasars (\qsoone\ and \qsotwo) at 
$z\approx 3.23$, with a projected separation of $\approx 1$ arcmin (or $\approx 500~\rm kpc$ at $z\approx 3$). 
The quasar pair has $\approx$ 20 hours of archival high resolution spectroscopy and another 20 hours of approved observations with UVES (P102A; PI V. D'Odorico). Ultra deep MUSE observations, combined with high-resolution absorption spectroscopy of these quasars that act as bright background sources, will enable a unique view of the connection between galaxies and the IGM and CGM simultaneously in absorption and emission, extending previous studies to lower-mass galaxies and lower SB limits.
In this first paper\footnote{Throughout, we adopt the following cosmological parameters: $H_0=70\, \rm{km \,s^{-1}\, Mpc^{-1}}$, $\Omega_\mathrm{M}=0.3$, and $\Omega_\Lambda=0.7$. In this $\Lambda$CDM cosmology, 1$^{\prime\prime}$ corresponds to 7.5 physical kpc.}, we present preliminary results based on the first $\approx 40$ hours  collected to date, and focus on the extended Ly$\alpha$ emission physically associated with the quasar pair.

\begin{figure}
\centering\includegraphics[width=0.85\columnwidth]{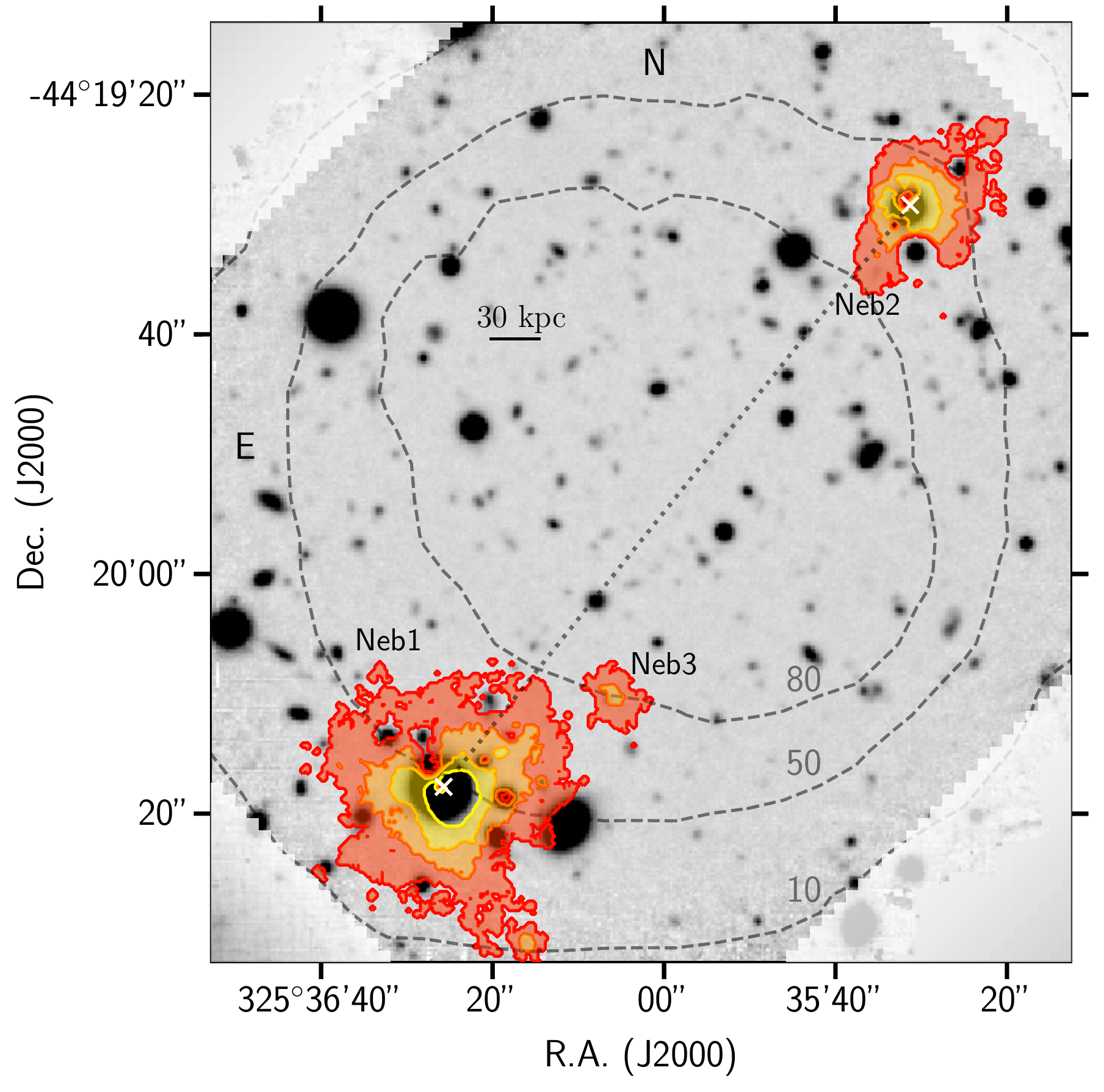}
\caption{Optimally extracted image of the MUDF (light grey) reconstructed from the 40-hour MUSE data cube.
Coloured contours represent the extended Ly$\alpha$ emission of the three nebulae detected within $\approx 2000~\rm km~s^{-1}$ of the two quasars, with contours of SB levels at 0.6, 3.2, 10, and 31.6$\times10^{-18}$~\sbline. The ``holes'' in the nebulae are caused by the subtraction of continuum sources. The dashed lines represent the contours at 10, 50 and 80 exposures/pixel from the combined mean optimally extracted image. The alignment and morphology of these nebulae offer a tantalising suggestion of the presence of large-scale filaments \rev{(orientation marked with a dotted line)} connecting the halos of QSO1 and QSO2 (marked with white crosses).}
\label{fig:fov}
\end{figure}

\begin{table*}
\centering
\caption{Global properties of the three \lya\ nebulae.} \label{tab:neb}
\begin{tabular}{ccccrrrc}
\hline
\hline
Object & \zn & $z_{\rm qso}$ & $m_{\rm r}^{\rm a}$ & FWHM$_{\rm Ly\alpha}^{\rm b}$ & $F_{\rm Ly\alpha}$ & $L_{\rm Ly\alpha}$ & Size$^{\rm c}$\\
 &      &    &  & km s$^{-1}$ & $10^{-18}~\rm erg~s^{-1}~cm^{-2}$ & $10^{43}~\rm erg~s^{-1}$ & kpc\\
\hline
Neb1 & 3.230 & 3.221$\pm$0.004 & 17.9$\pm$0.02 & 1124$\pm$23 & 748.9$\pm$3.5   & 6.81$\pm$0.03 & 140  \\
Neb2 & 3.229 & 3.229$\pm$0.003 & 20.5$\pm$0.03 & 1153$\pm$24 & 276.7$\pm$1.8   & 2.52$\pm$0.02 & 100  \\
Neb3 & 3.254 & --    & 27.1$\pm$0.20 & 513$\pm$25 & 35.5$\pm$0.6    & 0.32$\pm$0.01  & 35 \\
\hline
\end{tabular}
 \flushleft\begin{list}{}
 \item {\it Notes.} ${}^{\mathrm{a}}${ Continuum AB magnitudes at $\lambda=$6184.27\AA, convoluted with the SDSS $r^{*}$ filter and corrected for Galactic extinction, for the two quasars and the counterpart of Nebula 3. ${}^{\mathrm{b}}$ Full-width at half-maximum of the narrow Ly$\alpha$ emission line. ${}^{\mathrm{c}}$ Approximate size of the nebula at $6\times 10^{-19}$ erg~s$^{-1}$~cm$^{-2}$~arcsec$^{-2}$.}
 \end{list}
\end{table*}

\section{Data acquisition and analysis}\label{sec:dataredux}

\subsection{Observations and data reduction}

MUSE observations have been carried out using the Wide Field Mode with extended blue coverage\footnote{4650-9300 \AA\ with FWHM$\simeq$2.83\AA\ (170 km s$^{-1}$) at 5000~\AA.} during the nights from August 15$^{\rm th}$ 2017 up until July 18$^{\rm th}$ 2018. Data were acquired in a small mosaic composed of three overlapping positions to cover both the full extent of the pair separation ($62.2 \pm 0.1$ arcsec) and a region of $\approx 20$ arcsec around each quasar with the $\approx 1\times 1$ arcmin$^2$ field of view of MUSE (see Figure~\ref{fig:fov}). 
We collected a total of 102 exposures, 19 of which were obtained in August 2017 with integration times of $1200~\rm s$ each, plus 83 exposures lasting $1450~\rm s$ each. 
The total exposure time for the data presented in this work is therefore 39.7 hours. 
A small dither pattern with $\approx 1$ arcsec offsets and 10 degree incremental rotations has been applied to ensure optimal sampling of the field of view and to allow
for better control of systematic errors associated with the uneven response across the different spectrographs. 
Weather conditions during the runs were on \rev{average good, with clear sky} and sub-arcsecond seeing ($\approx 0.6-0.8~\rm arcsec$) for the majority of the exposures. The resulting image quality, 
 characterised by a full-width at half-maximum FWHM$=0.60\pm 0.01$ arcsec as measured across the field in the reconstructed white-light image, was enabled by the use of the GALACSI system \citep{2006NewAR..49..618S} that corrects the ground-layer turbulence across the entire field of view with four laser guide stars. 

Observations have been reduced with the standard ESO pipeline \citep[][v2.4.1]{Weilbacher2014} and combined in a single mosaic. We have further post-processed all the individual exposures using the \textsc{CubExtractor} package (v1.7; \citealt{cantalupo2019}) to improve the quality of the flat-field and the sky subtraction as detailed in \citet{borisova2016} and \citet{MF2016pristine,MF2017uvb}. We refer the readers to these works for further details.
After combining individual exposures in a final data cube using a 3$\sigma$ clipping mean, we rescaled the resulting variance (which is underestimated following the resampling of pixels in the final cubes) by a wavelength-dependent factor needed to match the effective pixel root-mean-square (rms) variation in each layer.

Throughout this work we also use near-infrared spectroscopy available from X-SHOOTER (ESO PID 085.A$-$0299) for \qsotwo\ (QSO2 hereafter), and that we collected using the FIRE spectrograph at the Magellan Telescope for \qsoone\ (QSO1 hereafter). The data have been reduced following the reduction techniques as discussed in e.g. \citet{lopez2016} and \citet{simcoe2011}. Additional details on the supporting spectroscopy available for the quasars and the adopted reduction techniques will be presented in a forthcoming publication. 
The quasar systemic redshifts are obtained from the combined constraints provided by the broad H$\beta$ and the narrow [O\,III] emission line doublet, finding $z_{\rm sys}=3.221\pm0.004$ ($\sigma_z\simeq280$ km s$^{-1}$) for QSO1 and $z_{\rm sys}=3.229\pm0.003$ ($\sigma_z\simeq250$ km s$^{-1}$) for QSO2.

\subsection{Analysis of the MUSE data cube}

Before proceeding to the analysis of the diffuse emission associated with the quasar pair, we post-processed the MUSE data in the following way \citep[see][]{borisova2016,MF2016pristine}. 
The point spread function (PSF) of the quasars and of the sourrounding stars was subtracted with the \texttt{CubePSFSub} method within \textsc{CubExtractor} \citep{cantalupo2019}. 
We then subtracted continuum sources in the field using the \texttt{CubeBKGSub} procedure and ran \textsc{CubExtractor} on the resulting cube to search for extended Ly$\alpha$ emission. To this end, we focused on a slice of the cube $\pm$40\AA\ ($\pm$9865 km s$^{-1}$) on either side of the \lya\ emission at the quasar redshift. 
In this slice, \textsc{CubExtractor} identified the Ly$\alpha$ extended emission around the quasars considering connected pixels that match the following criteria:  minimum volume of 2500 connected voxels above a signal-to-noise ratio of $S/N\geq2.5$; a minimum area of 500 spatial pixel$^2$ per detection (i.e. $\sim$35 kpc on a side); and a minimum number of spectral pixels per detection of 5. We smoothed the data by 3 pixels (0.6 arcsec) using a median filter in the spatial direction only.
With this selection we identified 3 extended Ly$\alpha$ nebulae, to a SB limit of $\approx 6\times 10^{-19}~$\sbline \rev{ (i.e.  $4\sigma$)}. \rev{For a comparison}, the pixel rms at this wavelength is $\approx 1.5\times 10^{-19}~$\sbcgs.

The results presented in this work are based on the final mean cube. Our findings do not depend on the methodology of combining the data, as we find the same result using either a mean or a median. Moreover, the detection of the nebulae and their global properties are independent on the selection criteria above, as verified by performing the extraction at different $S/N$ thresholds (i.e. 2, 3, and 5), or with different values of connected voxels (i.e. 1200, 3000) \rev{and  minimum area (i.e. 400 spatial pixel$^2$ or 600 spatial pixel$^2$).} 
We have also checked that the morphology of the nebulae does not depend on the observing strategy resulting in a non-uniform exposure coverage within the field. Indeed, the analysis of the first 6 hours taken with uniform exposure coverage yields a similar morphology once accounting for the different depth.

\section{Properties of the extended nebulae}

Our search for extended \lya\ emission has uncovered the presence of three nebulae with sizes $\gtrsim 35~\rm kpc$ on a side within $\approx 2000~\rm km~s^{-1}$ of the redshift of the two quasars. The SB maps of these nebulae, reconstructed following the method in \citet{borisova2016} by summing the line emission along the wavelength direction inside the three-dimentional segmentation map derived by \textsc{CubExtractor}, are shown in Figure~\ref{fig:fov}. 

\subsection{Morphological properties}
Nebula 1 is the most extended Ly$\alpha$ structure in the field, 
which was first identified by \citet[ID22]{2019MNRAS.482.3162A} with observations sensitive to $\approx 10^{-18}$~\sbline, and it is associated with the brighter quasar QSO1. With a diameter of $\approx 110 ~\rm kpc$ measured at $\approx 10^{-18}~\rm erg~s^{-1}~cm^{-2}~arcsec^{-2}$ and a full extent of $\approx 140 ~\rm kpc$ measured at $\approx 6\times 10^{-19}~\rm erg~s^{-1}~cm^{-2}~arcsec^{-2}$, this nebula has a total \lya\ luminosity of $(6.81\pm 0.03) \times 10^{43}~\rm erg~s^{-1}$. 
We also identify a second nebula, Nebula 2, that is physically associated with QSO2 and has a luminosity of $(2.52 \pm 0.02)\times 10^{43}~\rm erg~s^{-1}$ and size of $\approx 50~\rm kpc$ at $\approx 10^{-18}~\rm erg~s^{-1}~cm^{-2}~arcsec^{-2}$ ($\approx 100~\rm kpc$ at $\approx 6\times 10^{-19}~\rm erg~s^{-1}~cm^{-2}~arcsec^{-2}$). Finally, a third nebula (Nebula 3) is identified with a velocity offset of $\approx +1550~\rm km~s^{-1}$ compared to the two nebulae associated with the quasars. Smaller in size ($\approx 35~\rm kpc$ at $\approx 6\times 10^{-19}~\rm erg~s^{-1}~cm^{-2}~arcsec^{-2}$) and with luminosity of $(3.23\pm 0.07)\times 10^{42}~\rm erg~s^{-1}$, this nebula is associated with a faint continuum-detected source ($m_r =27.1 \pm 0.2$ mag), similarly to other mid-size nebulae \rev{(i.e. a few tens of kpc on a size)}  that are being uncovered by MUSE \rev{\citep[e.g.][ but see also \citealt{yang2009} for similar size nebulae powered by brighter objects]{fumagalli2017}}.
A summary of the properties of the nebulae is given in Table~\ref{tab:neb}.

A particularly striking feature of Fig.~\ref{fig:fov} is the relative alignment and morphology of this group of nebulae. Nebula 1 exhibits \rev{an asymmetrical shape}, with a prominent arm extending to the South-West and connecting to a compact emitter. \revs{Moreover, the nebula appears to be elongated along the North-West direction, towards QSO2}. Similarly, Nebula 2 is elongated along the same axis, with a clear extension pointing back towards Nebula 1. 

\revt{To quantify the elongation, we compute the asymmetry parameter $\alpha$ (the ratio between the semi-minor and semi-major axis) and the position angle describing the light distribution $\Phi$ as in \citet{2019MNRAS.482.3162A}, finding $\alpha \approx 0.80$ and $\Phi \approx -56.8$ deg for Nebula 1 and $\alpha \approx 0.91$ and $\Phi \approx -48.1$ deg for the Nebula 2. Within uncertainties (see \citealt{2019MNRAS.482.3162A} for details), the similar $\Phi$ values for both nebulae provide an additional indication of alignment along a common axis for these systems.}
\rev{Moreover, the probability of the nebulae to be randomly oriented can be estimated as the ratio between the subtended angle of the nebulae ($\approx 50$ degrees) and $2\pi$, $(50/360)^2$, which is only 2\%. The alignment of the nebulae thus likely reflects the presence of an underlying structure not currently detected.}

While not at the exact same redshift ($z\simeq 3.25$), Nebula 3 lies in between the two quasars, again aligned \rev{in projection} with the axis connecting QSO1 and QSO2.
Altogether, these features are suggestive of the presence of filamentary structures extending from the halos of the two host galaxies. These elongated nebulae, which appear to be a scaled-up version of what is inferred from the stacking of MUSE observations of pairs of \lya\ emitters \citep{gallego2018}, are thus hinting at the presence of a filaments connecting the nodes where halos form.
\rev{From a theoretical perspective, this picture is consistent with the structures
predicted in cosmological hydrodynamic simulations \citep[e.g.][]{fumagalli2011,vandevoort2012,rosdahl2012,schaye2015}. 
Our results, albeit on larger scale, are also comparable to the very extended \lya\ nebula discovered in a quasar pair at $z$=2.4 by \citet{cai2018}. With a projected separation of only  $\approx 80$ kpc, this pair is powering a common nebula with a size of $\approx 232$ kpc featuring an elongated morphology along the line connecting the two quasars in projection.}

Despite being clustered within a $\approx 500~\rm kpc$ region on a side, Nebula 1 and 2 lie within the parameter space defined by single quasars of comparable magnitude at these redshifts. In particular, they exhibit typical sizes and \lya\ luminosities (see e.g. fig. 3 in \citealt{borisova2016} and fig. 15 in \citealt{2019MNRAS.482.3162A}). This suggests that quasar nebulae trace halo gas within their host galaxies and that their clustered nature is a  consequence of the distribution of halos in CDM. Indeed, the expectation is that the ``isolated'' nebulae reported in the literature are themselves clustered with \lya\ nebulae that are just below the SB level reached by shallower observations in quasar fields \citep[see e.g.][]{wisotzki2018}.

\begin{figure}
\centering\includegraphics[width=0.8\columnwidth]{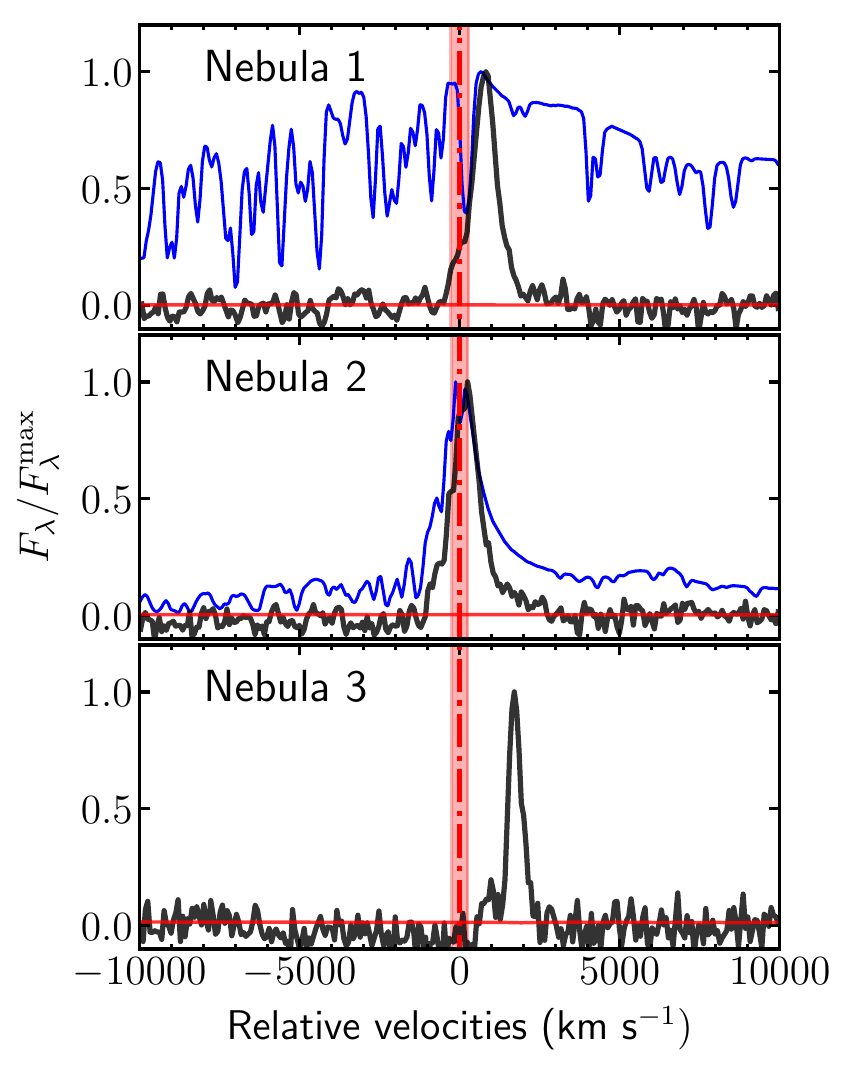}
\caption{MUSE spectra of the QSO1 and QSO2  (blue lines) and of the associated nebulae (black lines with related uncertainties in red) at the location of the Ly$\alpha$ emission. Velocities are relative to the systemic redshift of the quasars, whilst the average redshift of 3.23 is assumed as a reference for Nebula 3. Fluxes are normalized to their peak in this wavelength range. }\label{fig:spectra}
\end{figure}

\begin{figure*}
\includegraphics[width=0.69\columnwidth]{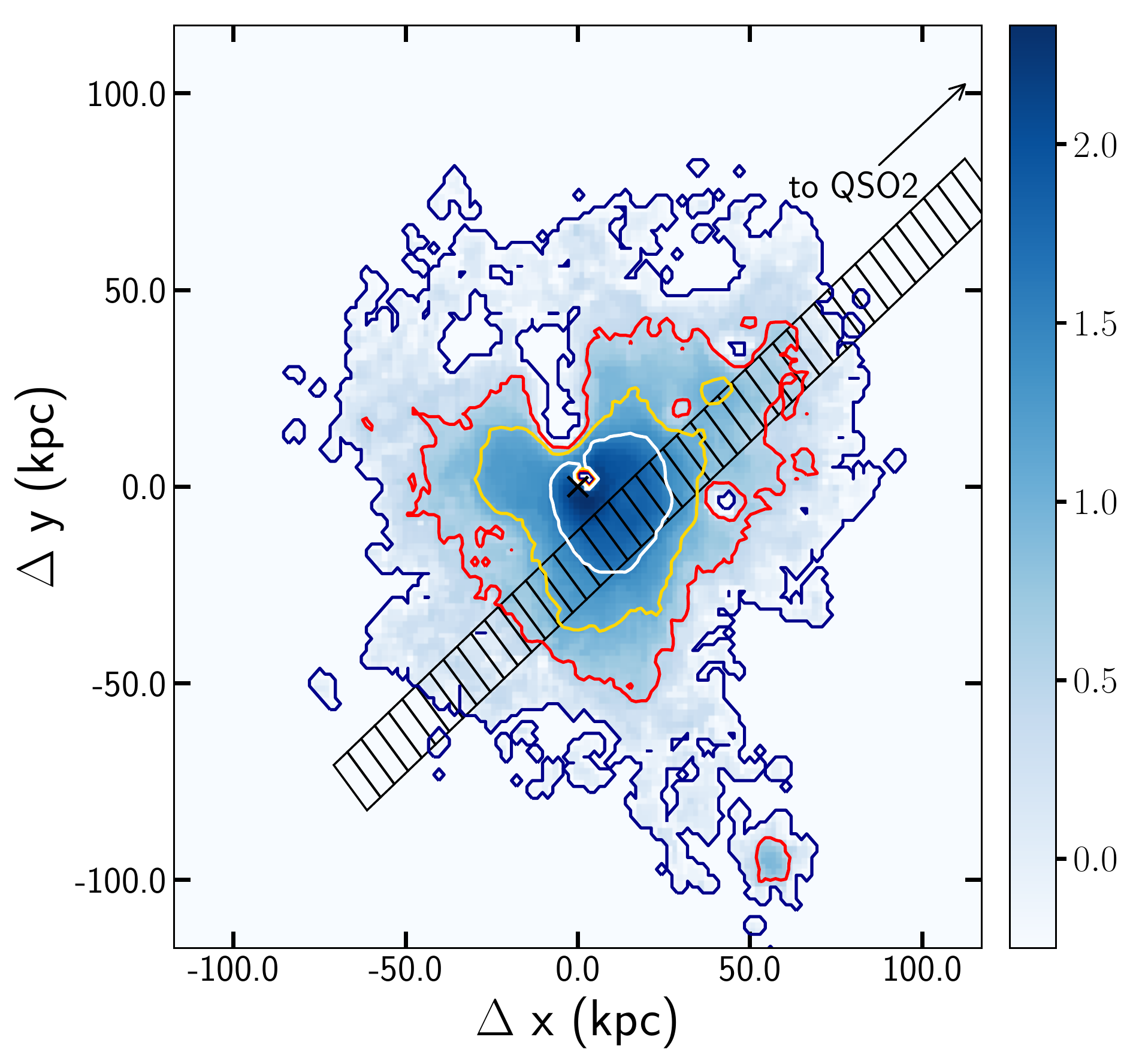}
\includegraphics[width=0.69\columnwidth]{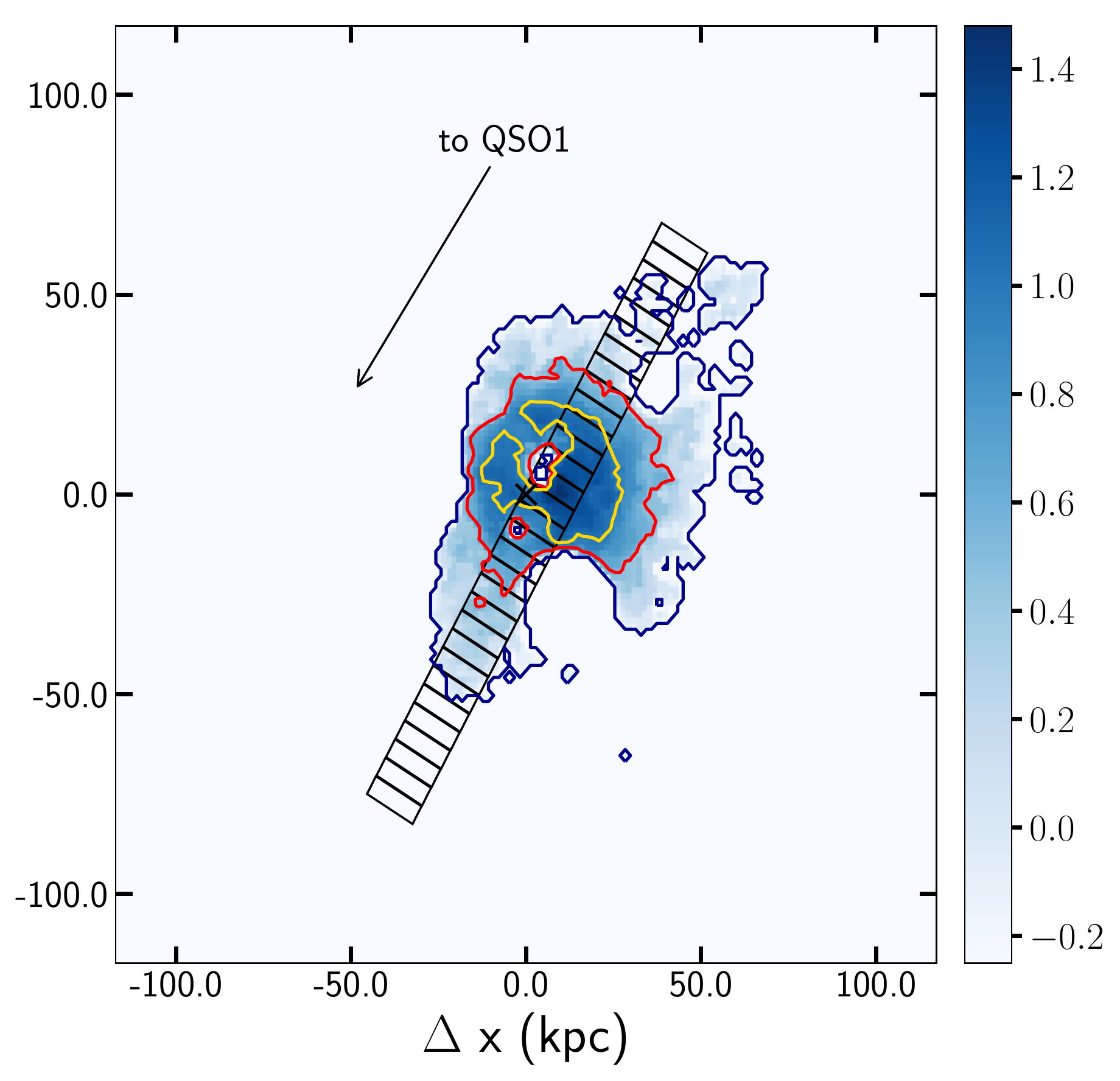}
\includegraphics[width=0.69\columnwidth]{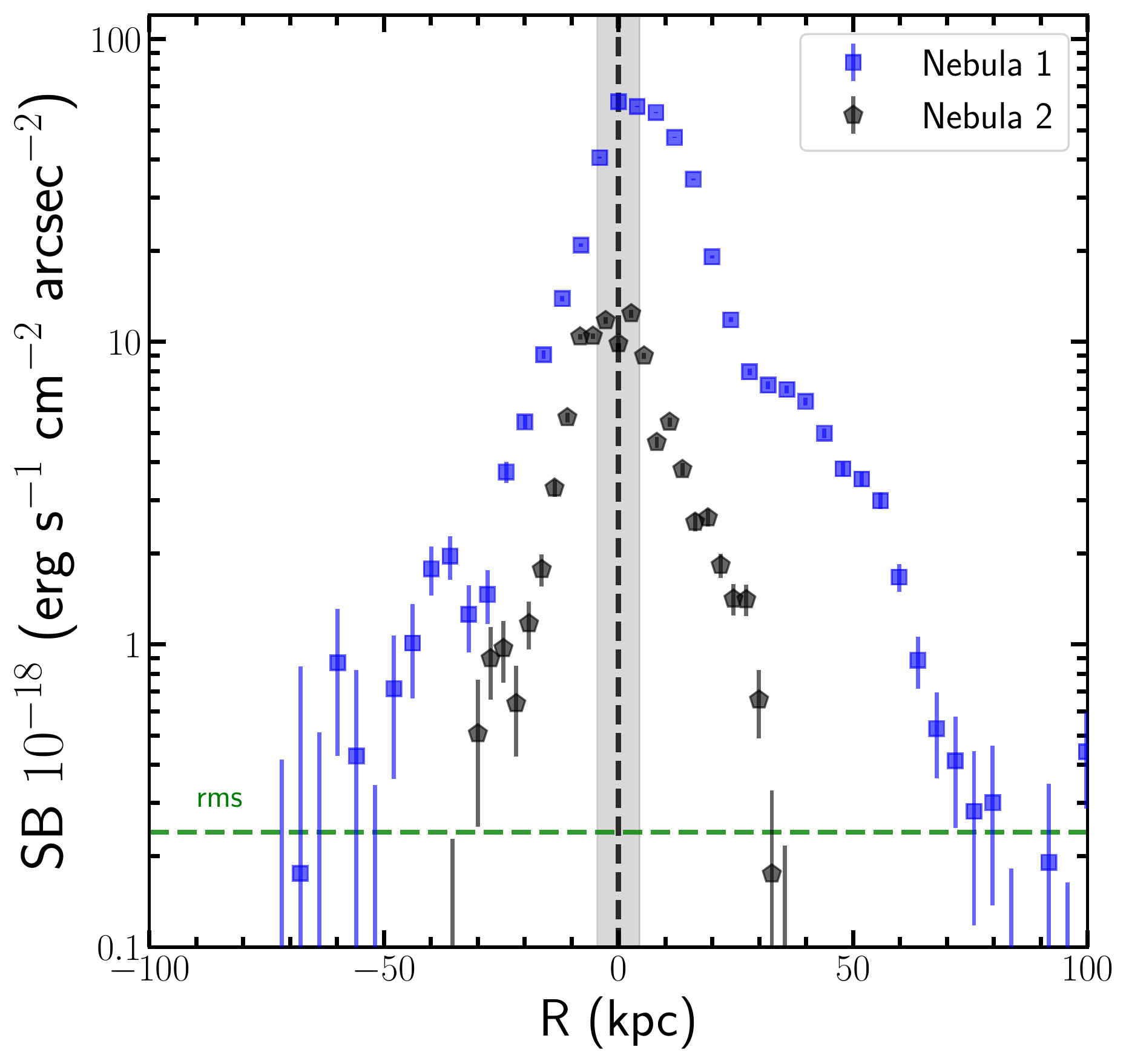}
 \caption{Ly$\alpha$ SB maps for Nebula 1 associated to QSO1 (left panel) and Nebula 2 associated to QSO2 (central panel) with contours at 0.6, 3.2, 10, and 31.6$\times10^{-18}~$\sbline\  in blue, red, yellow, and white, respectively. 
 The pseudo-long slits utilised to calculate the radial profiles are also shown. 
  {(\it Right)} SB radial profiles extracted along pseudo-slits in the direction of the expected filament connecting the two quasars computed from the narrow-band images. Also shown (green dashed line) is the empirical rms of $\simeq2.4\times10^{-19}$\sbline.
 The grey shaded area marks the average FWHM of the PSF estimated from stars within the field
 ($\approx 4.5$ kpc at $z\simeq3.23$). For clarity purpose, data below rms/2 are not shown.} 
\label{fig:SB}
\end{figure*}

\subsection{Spectral properties}

In Figure~\ref{fig:spectra}, we present the spectra extracted from the MUSE cube for QSO1 and QSO2 and the associated nebulae at the location of the Ly$\alpha$ emission, shown in velocity space
relative to the quasar systemic redshift. The spectra are extracted using pixels that spatially overlap with the {\sc CubExtractor} segmentation map, after we have masked out all the continuum sources detectable in the white-light image reconstructed from the MUSE cube. 
Fluxes are then normalised at the \lya\ flux peak emission. As found by \citet{borisova2016} and \citet{2019MNRAS.482.3162A}, the extended Ly$\alpha$ emission is characterized by a much narrower line profile than the broad Ly$\alpha$ emission of the quasars. This is expected for gas that traces material within the galaxy halo, outside the black hole sphere of influence. Nebula 1 also displays a small shift of $\approx 600~\rm km~s^{-1}$ with respect to the quasar redshift, an offset that has been also found in previous works (e.g. \citealt{2019MNRAS.482.3162A}). As deeper data becomes available, it will be possible to investigate in more detail the spatially-resolved kinematics of the nebulae, in connection with the quasar spectral properties.

The \lya\ in Nebula 3 is displaced by a velocity offset of $\approx +1550~\rm km~s^{-1}$ compared to Nebula 1 ($\approx +2200~\rm km~s^{-1}$ compared to the systemic redshift).
Considering the velocity offset as an upper limit to the Hubble flow velocity, Nebula 3 lies at $<5~\rm Mpc$ (proper) and hence it is likely to be associated with the same large scale structure hosting the quasars and within the region influenced by the quasars' radiation field (\citealt{cantalupo2014}).
With a line full width at half maximum of $513\pm 25\rm~km~s^{-1}$ this nebula is presumably powered by an obscured AGN, making it quite likely that several AGNs coexist within this structure, as seen for instance in the nebula discovered by \citet{hennawi2015}, \citet{2018MNRAS.473.3907A} and \citet[see also \citealt{2018A&A...610L...7H,2018A&A...614L...2H}]{cai2017}. This hypothesis can be tested with future X-ray observations.

\subsection{Surface brightness profiles}

As a final step, we derived the \lya\ SB profiles for Nebula 1 and Nebula 2 within the two pseudo-slits shown in Figure~\ref{fig:SB} along the direction of the expected filament connecting the two halos. For this calculation, we relied on narrow-band images of $\pm$15~\AA\ ($\pm$3,700 km s$^{-1}$) reconstructed around the \lya\ line centre of each nebula after subtracting continuum-detected sources from the data cube. 
The use of a narrow band image is preferable over the optimally extracted map as it conserves flux. 
To derive the profile, we average the flux in boxes of $\approx$ 34 pixels along the pseudo-slit, propagating the error accordingly. 
As the formal error does not account for the pixel covariance, we also calculate an empirical detection limit of $\approx 2.4\times10^{-19}~$\sbline\ on the SB from the distribution of fluxes calculated along $\approx 1000$ apertures of $\approx$ 34 pixels located randomly in the field and shown as a green dashed line in Figure~\ref{fig:SB}. 
This analysis highlights the different extents of the \lya\ emission around the quasars. In particular, Nebula 1 displays a clear excess of emission at $\simeq$50 kpc from the centre towards the direction of the expected filament. This excess, albeit less pronounced, may also be present in the Nebula 2.

When compared to the detection limit for line emission, it is evident that the SB profiles show no marked truncation radius (especially for the brighter Nebula 1, see right panel of Figure~\ref{fig:SB}) up to the current detection level. A reconstruction of the underlying gas distribution from the observed profile is not straightforward due to \lya\ radiative transfer effects and uncertainties in the mechanisms that power the emission \citep[e.g.][]{cantalupo2014,gronke2017}. However, at face value, the absence of a sharp decline in flux implies the presence of halo gas on scales larger than probed by current observations \citep{prochaska2013}. Upcoming deeper observations in the  MUDF will probe this gas distribution and the putative filament at larger distances from the  host galaxies.

\section{Conclusions}
We presented preliminary results from $\approx 40$ hours of observations in the MUSE Ultra Deep Field, a $1.2\times 1.4~\rm arcmin^2$ sky region at $21^h$:$42^m$:$24^s$ $-44^\circ$:$19^m$:$48^s$ that will be observed for a total of $\approx 200$ hours with MUSE, plus 90 orbits with the WFC3 G141 grism on board of HST. With the $\approx$40 hours of observations collected to date, we studied the extended \lya\ emission associated with a bright quasar pair at $z\approx 3.23$ with a projected separation of $500~\rm kpc$. Our primary findings are:  \\
$-$ We detected \rev{two extended} Ly$\alpha$ nebulae, physically associated with the two quasars (QSO1 and QSO2), with sizes of $\approx 140~\rm kpc$ and $\approx 50~\rm kpc$ measured at $6\times 10^{-19}~$\sbline. The two nebulae have \lya\ luminosities of $\approx 7\times10^{43}$ erg s$^{-1}$ and $2.5\times10^{43}$ erg s$^{-1}$, respectively. A third nebula is detected close to quasar QSO1 in projection, but at a velocity offset of $\approx 1550~\rm km~s^{-1}$. The three nebulae are believed to trace gas in halos clustered in the same large scale structure.  \\
$-$ Despite their clustered nature, the two quasar nebulae have global properties  (e.g. size and luminosity) in line with what found for isolated nebulae. However, their alignment and elongated morphology is suggestive of one or more gas filaments connecting the quasar host galaxies, as predicted by cosmological simulations. \\
$-$ After extracting SB profiles along pseudo-slits in the direction of the putative filament, we confirm that these nebulae present  asymmetric profiles along the line connecting the quasars. At the depth of our observations, we do not identify sharp edges, implying the presence of even more extended halo gas.

\rev{Upcoming observations in MUDF will enable a deeper view of the possible filament between the quasar pair, as well as of other structures that lie in the MUDF footprint at different redshifts.}

\vspace{-0.7cm}
\section*{Acknowledgements}
We thank the anonymous reviewer for useful comments.
EL is supported by a European Union COFUND/Durham Junior Research Fellowship (grant agreement no. 609412). We acknowledge support by the Science and Technology Facilities Council [grant number  ST/P000541/1] and from the European Research Council under the European Union's Horizon 2020 research and innovation programme (grant agreement nos. 717001 and 757535). This work is based on observations collected at ESO/VLT (ID 1100.A-0528). SC gratefully acknowledges support from Swiss National Science Foundation grant PP00P2\_163824. RJC was supported by a Royal Society University Research Fellowship.




\bibliographystyle{mnras}
\bibliography{bibl} 






\bsp	
\label{lastpage}
\end{document}